\documentclass[Physsubmission, Phys]{SciPost}

\hypersetup{
    colorlinks,
    linkcolor={red!50!black},
    citecolor={blue!50!black},
    urlcolor={blue!80!black}
}

\usepackage[bitstream-charter]{mathdesign}
\usepackage{siunitx}
\usepackage{physics}
\DeclareSymbolFont{usualmathcal}{OMS}{cmsy}{m}{n}
\DeclareSymbolFontAlphabet{\mathcal}{usualmathcal}
\DeclareSIUnit \parsec {pc}

\begin{document}

\begin{center}{\Large \textbf{
SUSY-QCD corrections to squark annihilation into gluons and light quarks\\
}}\end{center}

\begin{center}
Luca Paolo Wiggering\textsuperscript{1}
\end{center}

\begin{center}
{\bf 1} Institut für Theoretische Physik, Westfälische Wilhelms-Universität Münster 
\\
* luca.wiggering@uni-muenster.de
\end{center}

\begin{center}
\today
\end{center}


\definecolor{palegray}{gray}{0.95}
\begin{center}
\colorbox{palegray}{
  \begin{tabular}{rr}
  \begin{minipage}{0.1\textwidth}
    \includegraphics[width=30mm]{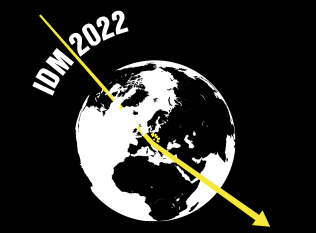}
  \end{minipage}
  &
  \begin{minipage}{0.85\textwidth}
    \begin{center}
    {\it 14th International Conference on Identification of Dark Matter}\\
    {\it Vienna, Austria, 18-22 July 2022} \\
    \doi{10.21468/SciPostPhysProc.?}\\
    \end{center}
  \end{minipage}
\end{tabular}
}
\end{center}

\section*{Abstract}
{\bf
We calculate the full one-loop SUSY-QCD corrections to stop annihilation into gluons and light quarks including the Sommerfeld enhancement effect and discuss their impact on the neutralino relic density in the MSSM. 
}


\section{Introduction}
\label{sec:intro}
Cosmological observations and simulations of structure formation provide strong evidence for the existence of cold dark matter (CDM). 
The analysis of the temperature anisotropies in the cosmic microwave background by the \emph{Planck} collaboration \cite{Planck:2018vyg} interpreted in the $\Lambda\text{CDM}$ model further allows to constrain the relic density of dark matter to the very precise range
\begin{align}
    \Omega_{\rm CDM}h^2 ~=~ 0.120 \pm 0.001
    \label{eq:omh2}
\end{align}
at $\SI{68}{\percent}$ confidence level, where $h$ denotes the present value of the Hubble parameter $H_0$ in units of $\SI{100}{\kilo\meter\per\second\per\mega\parsec}$. The assumption that dark matter consists of weakly interacting massive particles (WIMPs) does not only lead to the observed relic abundance via the WIMP miracle, but is also motivated by many BSM scenarios which predict the existence of new particles at the electroweak scale. One of these scenarios is given by the Minimal Supersymmetric Standard Model (MSSM), which provides under the assumption of $R$-parity conservation a stable WIMP in form of the lightest neutralino $\tilde{\chi}^0_1$. 

As supersymmetry only adds new particles, but leaves the interaction strengths unchanged at lowest order, it is natural to assume that today's neutralino abundance is produced thermally via the freeze-out mechanism, which is known to occur for other particles in the Standard Model (SM). The fact that every sparticle must eventually decay to the neutralino allows then to describe the evolution of the neutralino number density $n_\chi$ by the single Boltzmann equation 
\begin{equation}
    \dot{n}_\chi= -3 H n_\chi - \langle \sigma_{\text{eff}} v\rangle \left(n^2_\chi-(n_\chi^{\text{eq}})^2\right),
        \label{eq:boltzmann}
\end{equation}
where $n_\chi^{\text{eq}}$ denotes the equilibrium density and $H$ the Hubble rate\cite{Edsjo:1997bg}. The underlying particle physics model enters through the thermally averaged effective cross section 
\begin{equation}
    \langle \sigma_{\text{eff}} v\rangle = \sum_{i,j} \langle\sigma_{ij} v\rangle \frac{n^{\text{eq}}_i}{n_\chi^{\text{eq}}} \frac{n^{\text{eq}}_j}{n_\chi^{\text{eq}}},
    \label{eq:eff_XSec}
\end{equation}
which is a sum over all possible annihilation channels of sparticles into SM final states. The Boltzmann suppression factor 
\begin{equation}
    \frac{n^{\text{eq}}_i}{n_\chi^{\text{eq}}} \sim \exp\left(-\frac{m_i-m_\chi}{T}\right)
\end{equation}
gives the insight that besides pure neutralino annihilation, other (co)-annihilation channels can contribute significantly to the neutralino relic density if another sparticle is very close in mass to the neutralino. A theoretically well-motivated candidate for the next-to-lightest supersymmetric particle (NLSP) consistent with the observation of a SM-like $\SI{125}{\giga\electronvolt}$ Higgs boson is the lightest top squark $\tilde{t}_1$, as the mass splitting between the two chiral supersymmetric partners of a SM fermion is approximately proportional to the mass of the SM fermion, indicating the largest splitting for the partners of the heaviest fermion in the SM, i.e. the top quark. Another reason to consider light stops is that reconciling the increasing experimental limits on the neutralino mass with the observed DM density requires a mechanism to bring down the relic density which could be annihilation of the NSLP.  

The integration of the Boltzmann equation in Eq. \ref{eq:boltzmann} and the computation of the associated (co)-annihilation cross sections for generic models is nowadays a highly automatized process which is performed by public tools such as  \texttt{MicrOMEGAs} \cite{Belanger:2001fz}. However, these tools calculate the cross sections only at an (effective) tree level which is not a sufficient level of accuracy as the inclusion of next-to-leading order (NLO) corrections  can shift the predicted value of the dark matter relic density beyond the experimental uncertainty of the \emph{Planck} measurement. The impact of SUSY-QCD corrections were for example examined in Ref. \cite{Herrmann:2009wk} for the case of neutralino annihilation into heavy quarks. Thus, we extend the analysis of the impact of SUSY-QCD corrections on the neutralino relic density to the processes $\tilde{t}_1 \tilde{t}^\ast_1 \longrightarrow~ g g$ and $\tilde{t}_1 \tilde{t}^\ast_1 \longrightarrow~ q \bar{q}$ with $q$ denoting an effectively massless quark. These two processes have been implemented simultaneously into the software package \texttt{DM@NLO} as they have to be combined at NLO in order to obtain an infrared finite cross section.

\section{Computational details of the SUSY-QCD corrections}
\begin{figure}
    \centering
        \centering
        \includegraphics[width=0.17\textwidth]{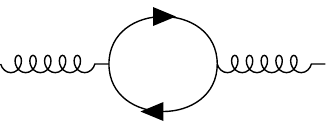} \ \ \ \
        \includegraphics[width=0.17\textwidth]{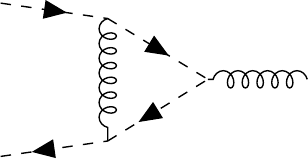} \ \ \ \
        \includegraphics[width=0.17\textwidth]{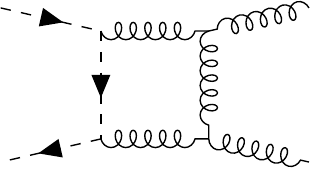} \ \ \ \
        \includegraphics[width=0.17\textwidth]{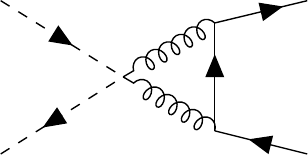} \ \ \ \
        \includegraphics[width=0.17\textwidth]{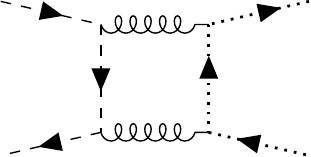} \ \ \ \
    \caption{A few example Feynman diagrams entering the virtual corrections.}
    \label{fig:diags_virtual}
\end{figure}

\begin{figure}
    \centering
        \includegraphics[width=0.14\textwidth]{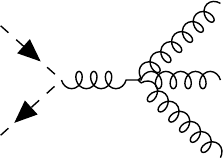} \ \ \
        \includegraphics[width=0.14\textwidth]{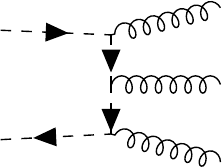} \ \ \
        \includegraphics[width=0.14\textwidth]{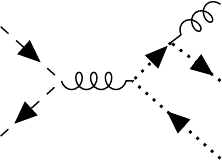} \ \ \ 
        \includegraphics[width=0.14\textwidth]{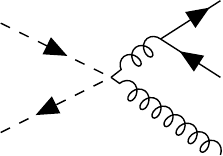} \ \ \
        \includegraphics[width=0.14\textwidth]{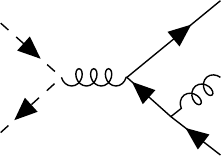} \ \ \
    \caption{A few example diagrams entering the real corrections.}
    \label{fig:diags_real}
\end{figure}

The NLO cross section 
\begin{equation}
    \sigma^{\text{NLO}} = \sigma^{\text{Tree}} + \Delta\sigma^{\text{NLO}}
\end{equation}
in the strong coupling $\alpha_s$  consists of the tree-level cross section $(\sigma v)^{\text{Tree}}$ and the NLO correction 
\begin{equation}
   \Delta\sigma^{\text{NLO}} = \int_2 \dd\sigma^{\text{V}}+\int_3\dd\sigma^{\text{R}} ,
\end{equation}
which is further subdivided into virtual $(\dd\sigma^{\text{V}})$ and real corrections $(\dd\sigma^{\text{R}})$. As we use the metric tensor $g^{\mu\nu}$ for the gluon polarization sum, the unphysical longitudinal polarizations of the external gluons are subtracted by including Faddeev-Popov ghosts as asymptotic states. In Figs. \ref{fig:diags_virtual} and \ref{fig:diags_real} a few example Feynman diagrams for both kinds of corrections are showcased where ghosts are represented as dotted lines and the arrows represent the ghost flow. 
The arising divergences occurring in the loops and within the three-particle phase space integration are regularized via the SUSY-preserving four-dimensional helicity scheme in $D=4-2\varepsilon$ dimensions \cite{Signer:2008va}. The ultraviolet divergences are then removed through a proper renormalization of fields, masses and couplings. In order to make the phase space integration of the real emission matrix element numerically accessible, the dipole subtraction method à la Catani-Seymour \cite{Catani:1996jh} is used, which has recently been extended to massive initial states in Ref. \cite{Harz:2022ipe}, so that it also applies to dark matter calculations. The central idea behind this subtraction method is to construct from the known singular behavior of (SUSY)-QCD amplitudes in the soft and collinear limit an auxiliary cross section $\dd\sigma^{\text{A}}$ that cancels the infrared divergences in the real-emission matrix element pointwise
and is at the same time simple enough so that it can be integrated analytically over the singular region allowing the cancellation of the infrared poles in the virtual part. The subtraction procedure can be summarized symbolically as  
\begin{equation}
    \Delta\sigma^{\text{NLO}}=\int_{3}\left[\dd{\sigma}^{\text{R}}_{\varepsilon=0}-\dd\sigma^{\text{A}}_{\varepsilon=0}\right]\\ +\int_{2}\left[\dd{\sigma}^{\text{V}}+\int_{1}\dd\sigma^{\text{A}}\right]_{\varepsilon=0}
\end{equation}
where the subscripts on the integrals refer to the dimensionality of the the phase space integrals.
An exchange of $n$ gluons between the incoming stop-antistop pair leads to a correction factor proportional to $(\alpha_s/v)^n$ for small relative velocities $v$ which means that these contributions become non-perturbative in the typical freeze-out regime and have to be resummed up to all orders. This is the well-known Sommerfeld enhancement effect \cite{Sommerfeld:1931qaf} and an analytical treatment is achieved via the framework of non-relativistic QCD, resulting in the cross section $\sigma^{\text{Som}}$. The full cross section $\sigma^{\text{Full}}$ is obtained by matching the fixed-order calculation to the resummed cross section. It is also possible to subtract the velocity enhanced part from virtual corrections giving the "pure" NLO correction $\sigma^{NLO}_v$.  For more computational details the reader is referred to our paper \cite{Klasen:2022ptb}.

\section{Impact on the annihilation cross section and the relic density }
\begin{figure}[h]
\centering
\includegraphics[width=0.48\textwidth]{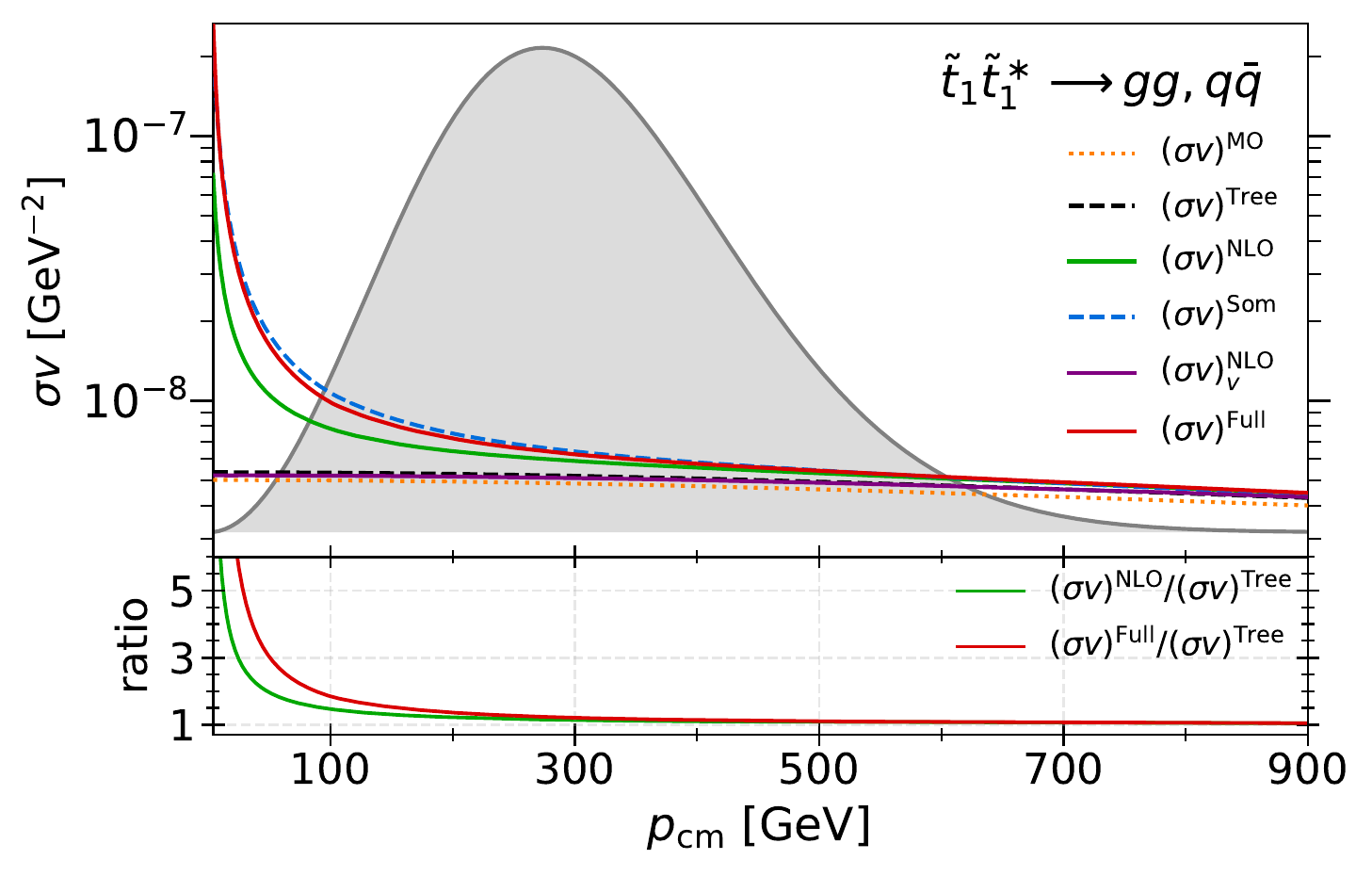}
\includegraphics[width=0.48\textwidth]{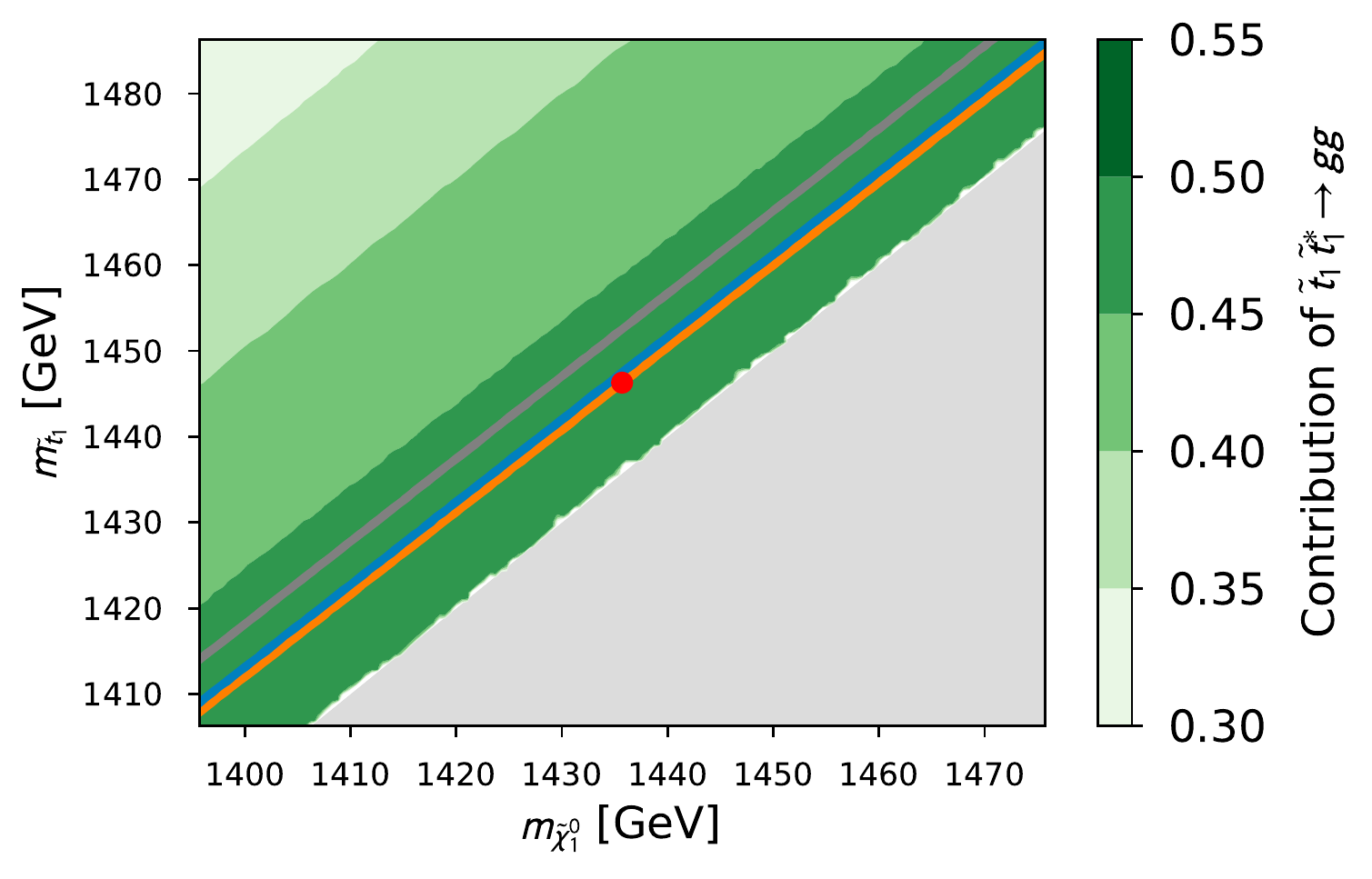}
\caption{Left: Annihilation cross section times velocity $\sigma v$ for stop-antistop annihilation into gluons and light quarks. The cross section obtained with \texttt{CalcHEP} is denoted as $\sigma^{\text{MO}}$. The gray region corresponds to the Boltzmann velocity distribution at the freeze-out temperature in arbitrary units. Right: Parameter region in the physical neutralino and stop mass plane that is consistent with the observed relic density at the $2\sigma$
confidence level. The orange band corresponds to the \texttt{MicrOMEGAs 2.4.1} calculation, the blue one to the \texttt{DM@NLO}
tree-level cross section and the gray band to the full corrected cross section. The red dot represents the chosen scenario. The different shades of green indicate the contribution of NSLP annihilation into gluons to the relic density.}
\label{fig:impact}
\end{figure}

In Fig. \ref{fig:impact} the impact of the radiative corrections on the annihilation cross section as well on the cosmologically preferred parameter region in the physical neutralino and stop mass plane is shown for a currently viable pMSSM-19 scenario with a mass difference $m_{\tilde{t}_1}-m_{\tilde{\chi}^0_1}=\SI{10.6}{\giga\electronvolt}$ between the bino-like neutralino and the stop. Consequently, stop annihilation into gluons is with $\SI{47}{\percent}$ the largest contribution to the relic density for this scenario followed by stop pair-annihilation into top quarks with $\SI{23}{\percent}$. The pure NLO corrections without the velocity enhanced part are below $\SI{\pm 3}{\percent}$ so that the full cross section is in very good approximation given by the Sommerfeld enhancement alone. However, the full corrections are still large enough to shift the relic density beyond the uncertainty of Planck data, resulting in an increased stop mass of $\SI{6.1}{\giga\electronvolt}$ compared to the \texttt{MicrOMEGAs} result to compensate the increased effective annihilation cross section. A more detailed discussion of the chosen scenario and the numerical results also for the additional inclusion of NLO corrections to stop-pair annihilation and neutralino-stop co-annihilation is available in \cite{Klasen:2022ptb}.

\section{Conclusion}
We examined the impact of NLO SUSY-QCD corrections including the Sommerfeld enhancement effect on the neutralino relic density focusing on contributions from stop annihilation into gluons and light quarks. We find that these corrections are important as they can shift the value of the relic density beyond the current experimental uncertainty. However, the overall NLO corrections are small so that the full cross section can be approximated by the Sommerfeld enhancement alone. We are also confident that this approximation extends to general dark matter models containing colored scalars.

\section*{Acknowledgements}
This work has been supported by the Deutsche Forschungsgemeinschaft (DFG, German Research Foundation) through the Research Training Group GRK 2149. L.W. thanks the organizers for the great conference.





\bibliography{SciPost_BiBTeX_File.bib}

\nolinenumbers

\end{document}